# Comparison of volume-selective z-shim and conventional EPI in fMRI studies using face stimuli

Hu Cheng
Srikanth Padmala
Rena Fukunaga

Department of Psychological and Brain Sciences
Indiana University
Bloomington, IN 47405

# Abstract

Single-shot gradient recalled echo planar imaging (EPI) is the primary tool for functional magnetic resonance imaging (fMRI). The image often suffers from signal drop near the air-tissue interface, such as the amygdala and regions of the orbitofrontal lobe. An effective way to correct for this type of artifact is by applying multi-shot EPI using different z-shimming. Unfortunately, the scanning efficiency is significantly lowered. More recently, a new technique called volume-selective z-shim was proposed to implement z-shim compensation to only specific slices with large susceptibility effects. The high imaging efficiency of volume selective z-shim makes it possible to substitute conventional EPI for whole brain studies. In this study two fMRI experiments were conducted to compare volume- selective z-shim and conventional EPI while subjects performed tasks on face stimuli. The comparison was focused on three brain regions: amygdala, hippocampus, and fusiform gyrus. Our results indicate that despite fewer volumes collected during the same amount of scan time, volume-selective z-shim showed statistically higher activation in brain regions susceptible to signal loss while minimal differences were observed elsewhere.

# INTRODUCTION

Over the years, Blood Oxygenation Level Dependent (BOLD) (1) based fMRI has become the most dominant imaging technique to explore brain functions. The BOLD contrast derives from differences in the T2* relaxation value that is induced by changes in the concentration of deoxyhemoglobin in the blood vessels. Usually a relatively long echo time is required to optimize BOLD sensitivity. However, a long echo time can cause signal loss in brain regions with large susceptibility effects. This signal loss is known to be mainly due to the intravoxel dephasing perpendicular to the slice plane, which is further worsened at high fields. The brain structures with the largest susceptibility artifacts are near the air-tissue interface, such as orbitofrontal lobe, amygdala, and inferior temporal lobes, which are brain regions often of considerable interest to researchers in cognitive neuroscience.

Researchers have made considerable attempts to reduce the through-plane signal loss for BOLD imaging among which include tailored RF pulses (2,3), multi-echo EPI imaging (4-6), dynamic shimming (7), passive shim (8-10), and special reconstruction approaches (11). While those methods have sufficed in many circumstances, one particular technique attracts considerable attention for its simplicity to solve the issues of field inhomogeneity by applying compensation gradients. Because the extra gradient is usually in the slice-selective direction, this technique has been coined "z-shim" (12-15). In the conventional z-shim method (12-14), several images are acquired using different gradient compensation levels, which then are appropriately combined to form one composite image. When using z-shim, there is an effective recovery of signal loss in the susceptible brain regions, however, this also comes at a great cost to the temporal resolution as a result of the extra acquisition of all the slices. This presents a challenge in dealing with a relatively long repetition time (TR) during whole brain imaging.

In order to resolve this problem, one proposal suggests that only one scan be performed with an adjusted refocusing gradient for each slice (16). The refocusing gradient is chosen to balance the signal gain from the susceptible regions and signal loss elsewhere. Hence, maximum z-shim effect is not achievable in this way. Another approach recommended for overcoming scan inefficiencies is called volume-selective z-shim (15). Here, the z-shim compensation is applied only to the selective volume with severe susceptibilities. By applying z-shim to only a selective region of the brain, this would help to minimize any increase in the overall scan time. In the method of Du et al., one image without z-shim compensation (referred as $NZS_c$ image) and one with z-shim compensation (referred as ZS image) were acquired for the selective slices within one TR, the two acquisitions are TR/2 apart. The effective TR for those images is TR/2, the flip angles are adjusted accordingly to reach maximum signal to noise ratio (SNR). For all other slices, only one image without compensation gradient (conventional EPI) was acquired. In comparison with conventional z-shim techniques, where different z-shim compensations are cycled through in successive acquisitions, volume selective z-shim is only a small deviation from conventional EPI, making it a practical substitute for conventional EPI in whole brain imaging studies. However, the z-shim effect is not as good as conventional z-shim because of the shorter effective TR for the selective slices. Some other factors may also compromise the benefit of z-shim. For instance, motion between the multiple acquisitions of the z-shimmed slice is hard to correct and will reduce the quality of the composite image; correlation of physiological noises from the multiple acquisitions can increase the noise in the composite image. How much an fMRI experiment is benefited in terms of statistical power by switching from conventional EPI to volume-selective z-shim is still unknown and a direct comparison between the two methods is desirable. The volume selective z-shim technique was applied in an fMRI experiment based on a reward-punishment task (15). Only a comparison of activation with and without z-shim was made in that study. The z-shim image was combined from both ZS and $NZS_c$ images, while the image for comparison is just the $NZS_c$ image with an effective repetition time of TR/2. It was not surprising that greater activation in the OFC was observed for z-shim image, simply due to the additional contribution of the image with z-shim compensation. A fair comparison should be between the two techniques with the same scanning time. Therefore, that comparison is not sufficient to convince fMRI researchers to choose volume selective z-shim over conventional EPI. On the other hand, although gradient compensation can effectively restore the signal in the susceptibility area, it may worsen the signal in other area. It is unclear how z-shim affects the activation in the region with negligible susceptibility that z-shim should not be applied. Furthermore, as fewer volumes are acquired for volume-selective z-shim, it is worthwhile to explore how this affects the fMRI sensitivity in the area where z-shim is not applied. To address all these questions, fMRI experiments were conducted using cognitive tasks that presented only neutral faces. This stimulus was chosen since previous studies have shown faces robustly activate several specific regions of the brain (17), including areas with high susceptibility artifacts such as amygdala (18). To our knowledge, this is the first study to fully investigate the efficacy of volume-selective z-shim compared to conventional EPI methods using face stimuli. Specifically, our study is intended to answer three questions. 1. How much is volume-selective z-shim superior to EPI in detecting brain activation in the susceptibility area? 2. How is BOLD signal affected in the selected volume for brain regions with little susceptibility effects? 3. In the non-selective volume, how is BOLD signal compromised by less scan efficiency? The results demonstrate volume-selective z-shim to be superior to conventional EPI methods; an overall advantage that is most evident during instances of considerably large through-plane field inhomogeneity.

## METHODS

### Pulse sequence

The volume selective z-shim pulse sequence was implemented on a Siemens 3T Trio system (Siemens Medical Systems, Erlangen, Germany). The custom sequence was written in the Siemens' pulse sequence development tool IDEA. The pulse sequence for this study followed similar protocols as Du and colleagues (15), including the use of only one additional scan for each z-shimmed slice. Some definitions were also inherited from Du et al., throughout this paper for consistency. The volume selected for z-shim is referred as Region-C. The volume not selected for z-shim is referred as Region-NC. For the images in Region-C, those acquired without z-shim compensation are named as $NZS_c$ images and those acquired with z-shim compensation are named as ZS images. There are some minor differences in our pulse sequence. Unlike the approach proposed by Du et al., z-shim was realized by altering the refocusing gradient moment. Assuming the refocusing gradient moment of a conventional EPI is $A_R$, a z-shim level is defined as the ratio of the offset refocusing gradient $\Delta A_R$ to $A_R$. The z-shim level can be positive or negative, depending on the polarity of the susceptibility-induced magnetic field gradient. If the z-shim level is 0, there will be an absence of compensation gradient and no z-shim effects. Since the refocusing gradient moment $A_R$ is linearly related to the inverse of the slice thickness, smaller z-shim level should be used for thinner slices. For the work in this paper, only one ZS image (z-shim level = 0) and one $NZS_c$ image (z-shim level > 0) were acquired for each slice in Region-C.

**Data acquisition**

All data were collected on a Siemens 3T Trio system. High resolution anatomical images were taken with a MP-RAGE sequence. Functional images with identical slices were acquired using conventional EPI and z-shim EPI. Subjects were presented with neutral face stimuli through a home built projection system. For z-shim EPI, TR/TE = 2500/28 ms, and a flip angle of 81° for slices without z-shim and 67° for slices with z-shim. The flip angles were calculated as the Ernst angle at the TR and TR/2, as proposed in (15). For conventional EPI, TR/TE = 2000/28 ms, and a 70° flip angle. This flip angle was chosen to maximize the BOLD effect by considering flow effects (19) and be consistent with the standard protocol in our institution. The signal difference between this flip angle and the Ernest angle (77°) is less than 1% for gray matter. According to a recent study, the effect of TR on the BOLD response is negligible in the range of 30° – 84° (20). For the EPI sequence used in the experiments, a z-shim level of 0.10 results in a compensation of gradient moment of 2.06 ms·mT/m at slice thickness of 4 mm and 2.47 ms·mT/m at slice thickness of 3.2 mm. To determine the appropriate z-shim level for individual subjects, prior to the functional scans, five test scans were performed using z-shim levels ranging from 0.02 to 0.12, adjusted from the gradient moment values used in (16). The z-shim level for each individual's functional scans was manually determined based on the maximized signal value in the amygdala.

Experiment 1

In the first experiment, we tested the main effects of z-shim on a large through-plane inhomogeneity. Six subjects (4 male, age range 24 – 39 years, mean age 27.5 years) participated in this experiment. All participants were in good health with no past history of neurological or psychological conditions. Each participant completed 6 counter-balanced runs that alternated between conventional EPI and z-shim EPI. Thirty-four axial slices were taken with a slice thickness of 4 mm with no gap. The functional blocks comprised of a 10 second neutral face presentation followed by a 10 second fixation. Each face was presented for 1500 ms followed by a 500 ms blank screen. The scan time for an individual run totaled 160 seconds. The subjects were told to memorize the faces for a later recognition test after the MRI scans. This experiment was specifically designed to target activation in the fusiform gyrus and also amygdala/hippocampus regions, which are known to be strongly modulated by face stimuli (17,18). Starting from the first slice, there were a total of 10 z-shimmed slices that covered amygdala and hippocampus to address question 1. The fusiform gyrus was included in the selective volume for z-shim to address question 2.

Experiment 2

In the second experiment we compared the effects of the z-shim methods while the EPI protocol is reasonably optimized for less susceptibility artifacts (21). Eleven subjects (6 male, age range 22 – 42 years, mean age 26.4 years) participated in this experiment. All participants were in good health with no past history of neurological or psychological conditions. Each participant completed 6 counter-balanced runs that alternated between conventional EPI and z-shim EPI.  Subjects were instructed to view neutral face stimuli and to identify the faces as either male or female. Thirty-four oblique slices 30˚ off the AC-PC line were taken with a slice thickness of 3.2 mm and no gap. The functional blocks comprised of a 20 second neutral face presentation followed by a 10 second fixation.  The male faces and female faces were presented randomly but counter-balanced in each block. Each face was presented for 1500 ms followed by a 500 ms blank screen. The scan time for individual runs totaled 240 seconds.  Starting from the sixth slice, there were a total of 9 z-shimmed slices that covered amygdala.  Although this experiment was again designed to target face-related activation in the fusiform gyrus and amygdala, by positioning the slices obliquely, the fusiform gyrus was not included in the selective volume for z-shim.  So in this experiment (unlike Experiment 1), we can also investigate how fMRI sensitivity is affected due to increased TR in the regions (for example fusiform gyrus)  where z-shim was not applied to address question 3.

**Data processing**

Image processing

The ZS images and NZS$_c$ images of the z-shimmed slices were combined using a square root of sum of squares (SSQ) approach to form composite images. A scaling factor calculated as the ratio of the mean signal of the two transition slices from Region-C to Region-NC was applied to the composite image to avoid any abrupt signal change between non-z-shimmed and z-shimmed slices.  A sample fMRI volume from Experiment 2 consisting of 34 slices, in which z-shim was applied to slice 6-14, is shown in Fig. 1A. The scanned images are grouped as $NZS_c$ slices, then slices in Region-NC, and ZS slices. The white lines indicate the $NZS_c$ slices (top) and ZS slices (bottom) with a z-shim level of 0.12. Fig. 1B shows the composite images, the images in the white box are combined from $NZS_c$ and ZS images in Fig. 1A.

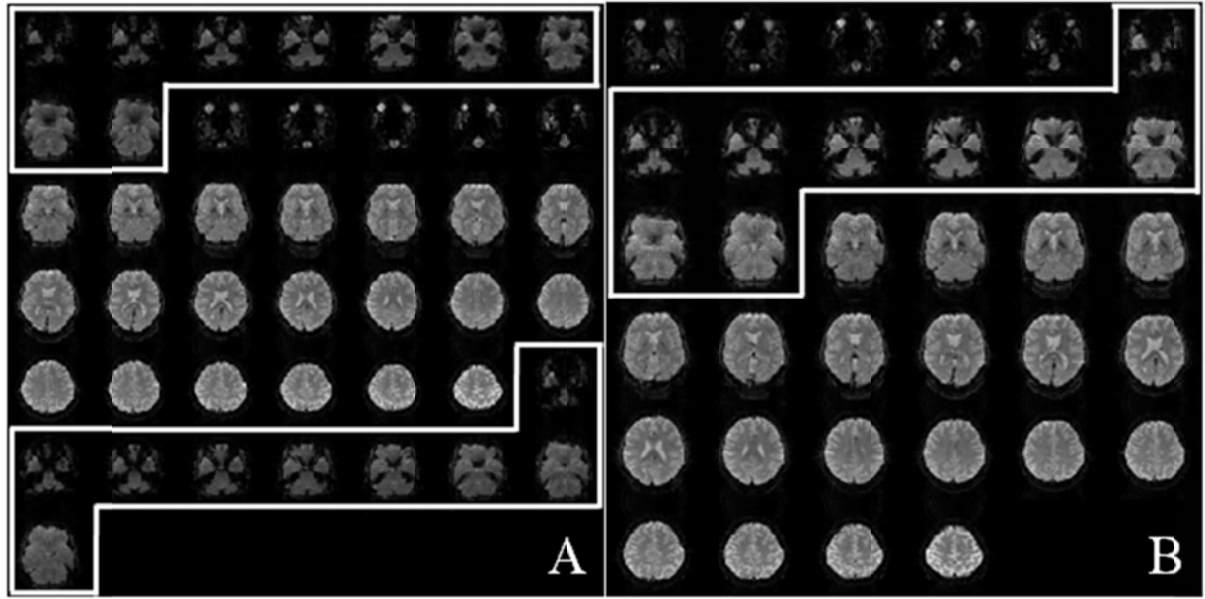

Fig. 1. Sample output of the acquired fMRI images using volume selective z-shim from Experiment 2. (A) Image consists of 34 total slices of which the beginning 9 images from slice 6-14 are the initially acquired without z-shim compensation ($NZS_c$ images), followed by 25 normal EPI slices, and then slice 6-14 acquired with z-shim compensation (ZS images). (B) A composite of the z-shimmed images (shown in white box) after the reordering and combining of the z-shimmed slices.

To quantitatively evaluate the signal recovery of the z-shim technique in the amygdala region, the mean signal in the amygdala was compared for conventional EPI and z-shim without scaling z-shimmed images. Instead of drawing the region of interest (ROI) for each subject, the amygdala was outlined from a template space and inversely mapped to the subject's individual image. For this procedure, the amygdala mask in the MNI T1 template was obtained using the WFUPickAtlas toolbox in SPM (22) downloaded from the website http://fmri.wfubmc.edu/cms/software#PickAtlas. The standard T1 template was normalized to the anatomical image of each individual subject, followed by a transformation applied to the amygdala mask. This mask was mapped to the functional image using the coregistration information between the anatomical image and functional image. Finally, the mean MRI signal was calculated by masking the amygdala onto the mean functional image. However, the TR for the z-shim scans was 2.5 seconds, while the TR for the conventional EPI scans was 2 seconds. Based on the use of unequal repetition times, a direct comparison between EPI and z-shim techniques is theoretically unsound. Assuming T1 of gray matter to be 1.33 seconds, the increase in the TR should result in a 7.5% signal gain at Ernst angles. Thus, before applying any comparisons, the signal based on z-shim EPI was multiplied by a correction factor of 0.93 (calculated using $1 / (1+0.075)$).

fMRI analysis

Both EPI and composite z-shim data were analyzed in SPM5 (Wellcome Department of Cognitive Neurology, London, UK) using identical procedures. After motion correction, the high-resolution anatomical image was coregistered and then resliced to the mean functional image. Followed by normalization of all functional images to the MNI standard T1 template, the warping was then applied to all functional images. After smoothing with an 8 mm Gaussian kernel, a GLM analysis was performed for each subject to obtain the single main contrast of face vs. baseline. High pass filter with a cutoff period of 128 seconds was used to remove slow varying drifts in the MR signal.

Two types of group analysis were performed for different purposes. To investigate the reliability of the brain activation at group level, a one-sample t-test was applied for the EPI and z-shim data separately by using the contrast images (con*.img) as input for the SPM5 second-level modeling. This analysis conducts a one-sample t-test by essentially taking the beta values from the fitted GLM conducted at the individual subject level. Then to compare the EPI and z-shim in the ability of detecting activation, whose difference is mainly a result of changes in SNR, a two-sample t-test was run on the t-score images (spmT*.img) from the two groups of individual GLM analysis. Assuming the subject's performance was stable across both EPI and Z-shim runs, the difference in t-score of the GLM analysis will be solely determined by the difference in BOLD contrast-to-noise ratio (CNR), which is closely related to changes in SNR. Therefore, t-score images are more appropriate for comparing the effect of SNR changes between EPI and Z-shim at the group level. Finally, in this study, we restricted our group analysis to the fusiform gyrus, amygdala and hippocampus ROIs in Experiment 1 and to the fusiform gyrus and amygdala ROIs in Experiment 2. This has been done to explicitly mitigate the multiple comparison problems and to focus on face related activation in both experiments and also the memory related activation in Experiment 1. For the first group analysis in these a priori defined regions, we used P < 0.005 (uncorrected) as statistical threshold and for the second group analysis comparing EPI and z-shim directly, we used P < 0.05 (uncorrected) as statistical threshold.

The SSQ approach does not change the noise amplitude if the noises from different z-shim acquisitions are not correlated (23). The noise may no longer stay constant in the presence of physiological noise which correlates both temporally and spatially (24). The temporal noise was compared for EPI and z-shim images (without applying a scaling factor) in Experiment 2 after motion correction and normalized to the standard MNI template. The temporal noise was computed as the standard deviation of the voxel-wise time course after a quadratic detrending. A mean noise map was generated for each subject by averaging the temporal noises from three runs. Thus, two sets of mean temporal noise maps were obtained for EPI and z-shim. A two sample t-test was conducted on the two data sets to investigate the difference in temporal noise of the two imaging methods. We used P < 0.05 (uncorrected) with a cluster size of > 10 voxels as statistical threshold. The activated voxels from previous one-sample t-test group analysis for z-shim were excluded in this statistical analysis to reduce contamination from BOLD signal change.

## RESULTS

Table 1 lists the signal strength in the amygdala using both z-shim and EPI. The signal values for z-shim have been corrected to take into account the effect due to the different TRs. For all functional scans, the z-shim approach produced higher signals in the amygdala. Since the optimal z-shim level differed among subjects, the results also indicated variability in the degree of signal enhancement. Experiment 1 demonstrated a larger signal enhancement due to the larger slice thickness, in which the susceptibility reduced gradient (15,25) lead to greater signal loss as a result of dephasing. Furthermore, the orientation was not chosen for minimizing through-plane homogeneity. Hence, the z-shim level for this experiment was slightly bigger. At the assumption that the SSQ combination does not alter noise level, these numbers/values indicated that the signal increase may have improved the SNR in the amygdala, from 6% to 23% (mean 15.2%) for Experiment 1 and 7% to 17% (mean 11.4%) for Experiment 2.

Table 1. A description of the signal strengths found in the amygdala across all subjects in Experiments 1 and 2 using both imaging sequences. The listed signals for z-shim have been corrected to take into account the effects of the TR difference between z-shim and conventional EPI. While there is higher signal intensity in the amygdala for the z-shimmed images, the results also indicate considerable variability in the optimal z-shim levels, as well as the signal enhancements found across all subjects.

| Subject | Experiment 1 | | Experiment 2 | |
|---|---|---|---|---|
| | z-shim signal strength (z-shim level) | EPI signal strength | z-shim signal strength (z-shim level) | EPI signal strength |
| 1 | 252 (0.10) | 238 | 309 (0.10) | 283 |
| 2 | 248 (0.10) | 208 | 232 (0.08) | 214 |
| 3 | 254 (0.10) | 219 | 294 (0.12) | 269 |
| 4 | 286 (0.08) | 248 | 254 (0.12) | 233 |
| 5 | 254 (0.08) | 207 | 429 (0.02) | 400 |
| 6 | 282 (0.08) | 252 | 421 (0.05) | 380 |
| 7 | | | 381 (0.04) | 342 |
| 8 | | | 428 (0.02) | 376 |
| 9 | | | 418 (0.04) | 363 |
| 10 | | | 405 (0.08) | 344 |
| 11 | | | 447 (0.04) | 393 |

Figure 2 shows the group activation maps from Experiment 1. As previously mentioned, the results are masked with three ROIs, i.e., the amygdala, hippocampus and fusiform gyrus. The activation maps using conventional EPI and z-shim methods are grouped in the top panel and the bottom panel, respectively. Activation in the amygdala was well detected using z-shim (Fig. 2D) but not conventional EPI (Fig. 2A). Strong brain activity was observed in the fusiform gyrus using both z-shim and conventional EPI (Fig. 2C and 2F). This is most likely due to the processing of the facial stimuli (17). However, the memory-related activation of the hippocampus is only observed in the z-shim method (Fig. 2E). Based on the t-test maps comparing EPI and z-shim techniques shown in Figures 3, the z-shim approach showed statistically stronger activation mainly in the amygdala and hippocampal region, while no differences were detected in the fusiform gyrus. These results suggest that z-shim may have helped to improve activation in the amygdala and hippocampus areas without causing any significant signal loss in other well-shimmed region, such as the fusiform gyrus.

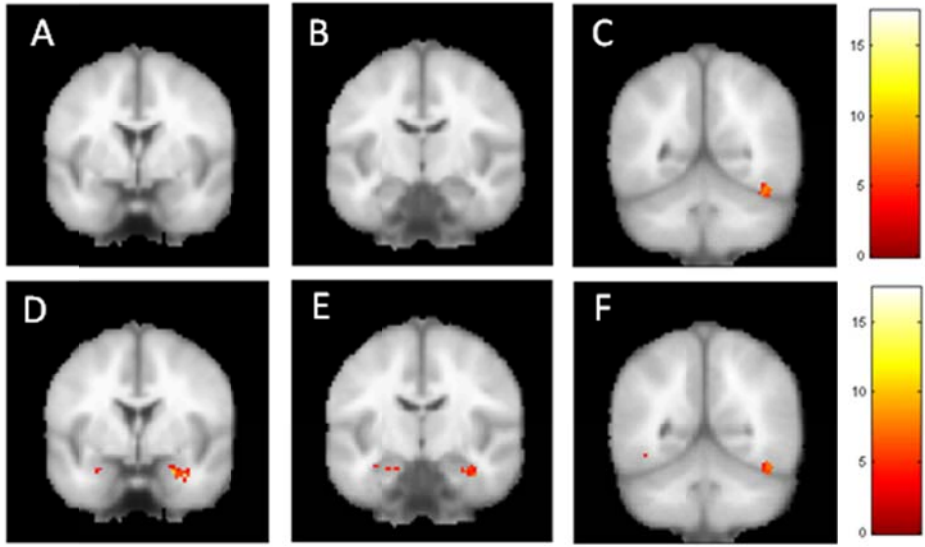

Fig. 2. Activation maps from Experiment 1 based on 6 subjects in one sample t-test group analysis at p < 0.005 (uncorrected). The maps are displayed in the coronal planes of the MNI standard T1 weighted image. Top panels are the results using conventional EPI in the region of (A) amygdala (y = 0 mm), (B) hippocampus (y = -10 mm), and (C) fusiform gyrus (y = -54 mm). Bottom panels show the results in the same brain area using z-shim: (D) amygdala (y = 0 mm), (E) hippocampus (y = -10 mm), and (F) fusiform gyrus (y = -54 mm).

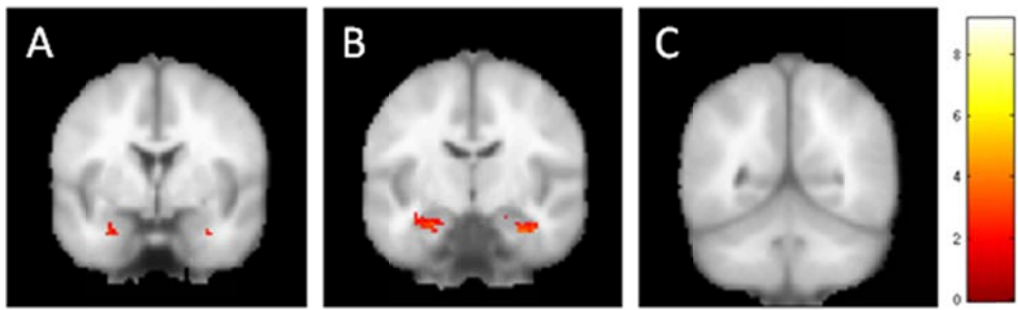

Fig. 3. Comparison of z-shim and conventional EPI using two-sample t-test on the t scores from the individual level analysis, thresholded at p < 0.05 (uncorrected). The maps show the statistical result of z-shim > EPI in the coronal planes of the MNI standard T1 weighted image for (A) amygdala (y = 0 mm), (B) hippocampus (y = -10 mm), and (C) fusiform gyrus (y = -54 mm).

Figure 4 shows group analysis activation maps from Experiment 2 from conventional EPI (4A-B) and z-shim (4C-D) runs. These results are focused on the ROIs of the amygdala and fusiform gyrus. Similar to results from Experiment 1, both the fusiform and amygdala were activated as a result of using face stimuli. Unlike Experiment 1, the z-shim approach did not show stronger statistical power than the EPI method in the amygdala from the group analysis of t-test maps. The p-value of 0.05 (uncorrected) indicated no significant difference between the two methods in either the amygdala or fusiform gyrus. Therefore, although when compared to EPI methods, z-shim method may not have been as efficient in the fusiform gyrus area because of undesired z-shim gradient and less efficiency, there were no significant differences in terms of functional activation.

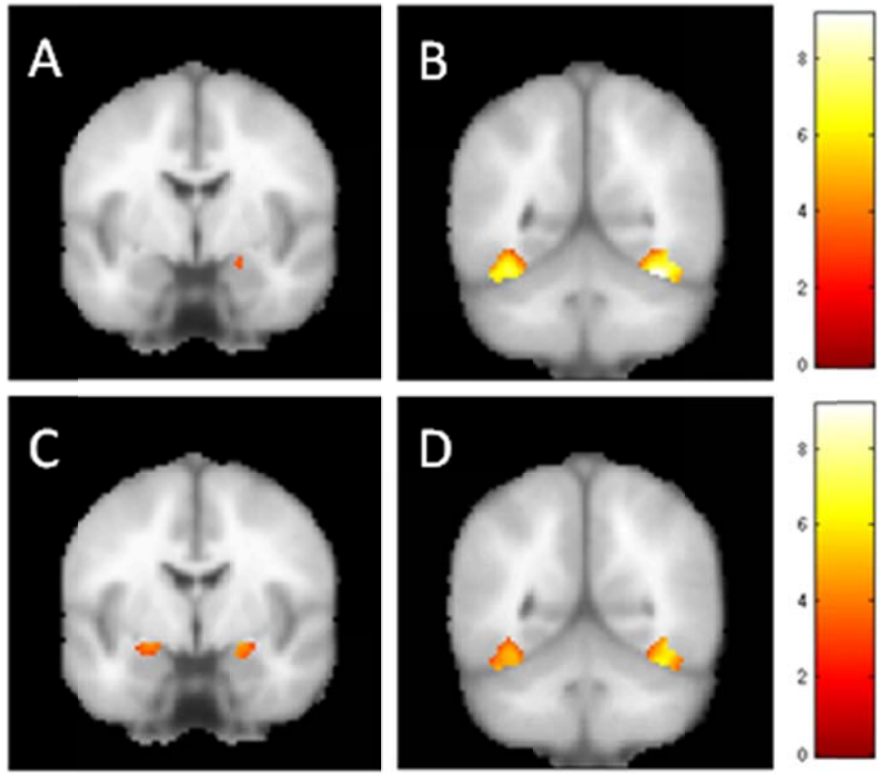

Fig. 4. Activation maps from Experiment 2 based on 11 subjects in one sample t-test group analysis at p < 0.005 (uncorrected). The maps are displayed in the coronal planes of the MNI standard T1 weighted image. Top panels are the results using conventional EPI in the region of (A) amygdala (y = -4 mm) and (B) fusiform gyrus (y = -54 mm). Bottom panels show the results in the same brain area using z-shim: (C) amygdala (y = -4 mm) and (D) fusiform gyrus (y = -54 mm).

Figure 5 shows the results of group comparison of temporal noise between conventional EPI and volume selective z-shim from a two-sample t-test. Fig. 5A and 5B are z-shim > EPI and EPI > z-shim respectively. At p-value of 0.05, no significant differences were observed in most area of the brain whether in Region-C or Region-NC. The few scattered clusters showing higher noise for EPI or z-shim are mainly located near the brain stem or in the cerebellum.

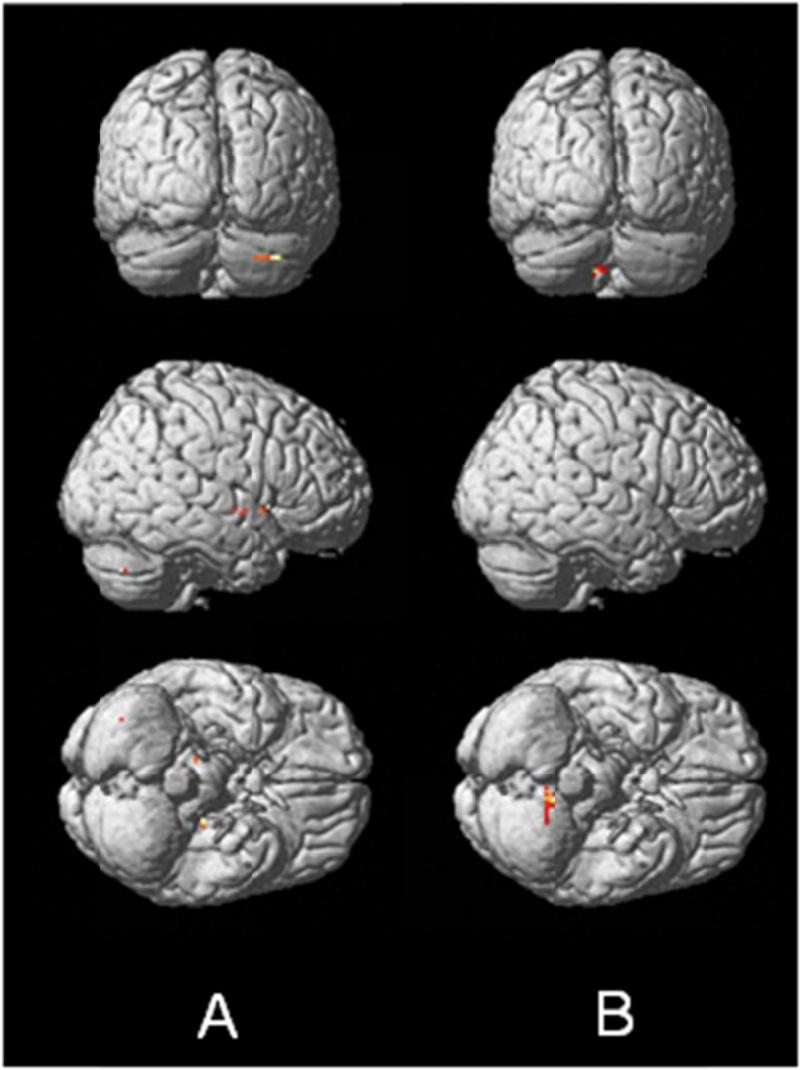

Fig. 5. Comparison of temporal noise between z-shim and conventional EPI using two-sample t-test on the mean noise map of individual subjects, thresholded at p < 0.05 (uncorrected) and 10 voxel cluster extent. The maps show the statistical results overlaid on a volume rendered brain for (A) z-shim > EPI and (B) EPI > z-shim.

## DISCUSSION

Without significant loss of scan efficiency, volume selective z-shim is very likely to be an alternative of conventional EPI in whole brain imaging. In addressing the three questions proposed in the introduction, our functional imaging study using face stimuli demonstrated the advantage of volume-selective z-shim EPI over conventional EPI, particularly in the amygdala and hippocampal regions. In those areas, volume-selective z-shim produced higher activation maps at the group level. The advantage of volume selective z-shim is most prominent with thicker slices and axial orientation, in which the intra-voxel dephasing effect is larger compared to thinner slices and oblique orientation. Our results showed only a small difference in the activation for both methods in the well-shimmed areas, such as the fusiform gyrus, whether they are selected for z-shim or not.

Volume selective z-shim is theorized as a superior method apart from conventional EPI in the selective volume for the following two main reasons: 1. The repetitions in acquiring the same volume and 2. The ability to effectively compensate field inhomogeneity by adjusting the z-shim parameters. The fMRI detection power at a given P value ($P$) is related to the number of scans and temporal SNR (26) by the following equation:

$$erfc^{-1}(P) = \sqrt{\frac{N}{8}}(TSNR)(eff), \tag{1}$$

where *TSNR* is the temporal SNR, *eff* is the effect size, $N$ is the number of scans, and *erfc* is the complimentary error function. When considering the same time length, there are fewer images with TR 2.5 seconds than with TR 2 seconds. The ratio is $\sqrt{2/2.5} = 0.894$. For the z-shim slices, if the z-shim level is set to 0 with only two acquisitions for the z-shim slices, the effective TR will be around 1.25 seconds. The MR signal can be modeled as a function of TR, T1 and flip angle as shown in equation 2

$$S \propto \frac{1-\exp(-TR/T1)}{1-\cos\theta \cdot \exp(-TR/T1)}\sin\theta \tag{2}$$

The signal of the composite image is then 17.3% higher than that from conventional EPI with 2 second TR, assuming a T1 value of 1330 ms at 3T. Thus the ultimate gain in statistical power is (1.173×.894 – 1) ~ 5%. This implies that even without applying a z-shim compensation gradient, the pulse sequence can be useful in improving the statistical power assuming the physiological noise remains the same in the composite image. The signal intensity can be further increased in the susceptibility regions by changing the z-shim levels. By taking the correction factor of 0.93 previously used to adjust for TR difference, a 17% signal increase from TR = 2.5 s corresponds to a signal increase of (1.17×0.93 – 1) ~ 8.8% for TR = 2 s. Our data in table 1 shows that the signal gain in the amygdala averaged 15.2% for Experiment 1 and 11.4% for Experiment 2, values both higher than the 8.8% gain without z-shim. In Experiment 1, the signal enhancement made a significant impact on the amygdala. Figure 3A shows the main contrast image between z-shim and EPI masked by the amygdala at a threshold of $p < 0.05$, uncorrected. In Experiment 2, however, the signal gain was not sufficient to impact any susceptible brain regions. These findings indicate that by optimizing the scanning protocol (e.g., decreasing slice thickness, changing the orientation), z-shim did not have a significant advantage over EPI in terms of the statistical power (t-test). While this may suggest minimal opportunities to make further improvements using z-shim methods, the increase in brain activation found at the group-level analysis by comparing Fig. 4A and Fig. 4C indicates that z-shim may help in other ways, such as reducing inter-subject variation.

For the slices without z-shim (in Region-NC), the intensity of a TR of 2.5 seconds vs. a TR of 2 seconds at corresponding Ernst angles is 1.075, therefore, there is only about ($1.075 \times .894 - 1$) ~ 4% penalty in ultimate statistical power for the slices without z-shim when shifting from a TR of 2 seconds to 2.5 seconds. The result from Experiment 2 suggested that despite fewer volumes collected during the same amount of scan time, there was no significant difference of BOLD activation in non-selective volume (Region-NC) between the two methods. For the slices with z-shim (in Region-C) but with minimal or no susceptibility effects, there is a mixed effect of signal enhancement from double acquisitions and signal drop from z-shim. The result from Experiment 1 showed that for the brain region with negligible susceptibility artifacts such as the fusiform gyrus, z-shim had little effect on the statistical power in this region (Region-C).

As indicated in equation 1, the statistical power is determined by the temporal SNR rather than the signal itself. The temporal noise was composed of both scanner noise and physiological noise. The scanner noise is considered to be the random thermal noise. The physiological noise is not white in nature. Depending on the sampling rate and the temporal correlation of the physiological noise, summation of physiological noise results in more or less noise, which may affect the slices in Region-C. For the slices in Region-NC, the MR signal was larger for z-shim because of the longer TR. Since the physiological noise is positively related to the signal strength (27), a higher temporal noise was expected in the non-z-shimmed slices. A close look at the temporal noise may further reveal its contribution to TSNR changes in z-shimmed slices. Despite those factors mentioned above, Fig. 5 demonstrated that the temporal noise was not significantly different for conventional EPI and volume selective z-shim in our study. Therefore, the signal intensities at amygdala listed in Table 1 are good indicators of the temporal SNR in that region.

In summary, we conducted two fMRI experiments to compare volume selective z-shim and conventional EPI techniques on whole-brain fMRI applications. For the areas with large susceptibility including the amygdala and hippocampus, more activation was discovered at the group level for volume-selective z-shim. For the area without susceptibility problem such as the fusiform gyrus, only a small difference was observed in brain activation for both methods, whether they are in the z-shimmed slices or not. The results indicate a statistical advantage in implementing z-shim techniques over conventional EPI sequences in brain regions known to suffer from large susceptibilities near the air-tissue interface.

## REFERENCE

1. Ogawa S, Lee TM, Kay AR, Tank DW. Brain magnetic resonance imaging with contrast dependent on blood oxygenation. Proc Natl Acad Sci USA 1990;87:9868-9872.
2. Chen N-k, Wyrwicz AM. Removal of Intravoxel Dephasing Artifact in Gradient-Echo Images Using a Field-Map Based RF Refocusing Technique. Mag Reson Med 1999;42:807-812.
3. Stenger AV, Boada FE, Noll DC. Three-Dimensional Tailored RF Pulses for the Reduction of Susceptibility Artifacts in $T$*2-Weighted Functional MRI. Mag Reson Med 2000;44:525-531.
4. Posse S, Shen Z, Kiselev V, Kemna LJ. Single-shot T2* mapping with 3d compensation of local susceptibility gradients in multiple regions. NeuroImage 2003;18:390-400.
5. Posse S, Wiese S, Gembris D, et al. Enhancement of BOLD-Contrast Sensitivity by Single-Shot Multi-Echo Functional MR Imaging. Mag Reson Med 1999;42:87-97.
6. Weiskopf N, Klose U, Birbaumer N, Mathiak K. Single-shot compensation of image distortions and BOLD contrast optimization using multi-echo EPI for realtime fMRI. NeuroImage 2005;24:1068-1079.
7. Koch KM, McIntyre S, Nixon TW, Rothman DL, de Graaf RA. Dynamic shim updating on the human brain. J Magn Reson 2006;180:286-296.

8. Wilson J, L., Jenkinson M, Jezzard P. Protocol to determine the optimal intraoral passive shim for minimisation of susceptibility artifact in human inferior frontal cortex NeuroImage 2003;19:1802-1811.
9. Wilson J, L., Jenkinson M, Jezzard P. Optimization of Static Field Homogeneity in Human Brain Using Diamagnetic Passive Shims. Mag Reson Med 2002;48:906-914.
10. Hsu J-J, Glover GH. Mitigation of Susceptibility-Induced Signal Loss in Neuroimaging Using Localized Shim Coils. Mag Reson Med 2005;53:243-248.
11. Liu G, Ogawa S. EPI Image Reconstruction With Correction of Distortion and Signal Losses. J Magn Reson Imaging 2006;24:683-689.
12. Constable RT. Functional MR imaging using gradient-echo echo-planar imaging in the presence of large static field inhomogeneities. J Magn Reson Imaging 1995;5:746-752.
13. Glover GH. 3D z-shim method for reduction of susceptibility effects in BOLD fMRI. Mag Reson Med 1999;42:290-299.
14. Yang QX, Dardzinski BJ, Li S, Eslinger PJ, Smith MB. Multi-gradient echo with susceptibility inhomogeneity compensation (MGESIC): demonstration of fMRI in the olfactory cortex at 3.0 T. Mag Reson Med 1997;37:331-335.
15. Du YP, Dalwani M, Wylie K, Claus E, Tregellas JR. Reducing Susceptibility Artifacts in fMRI Using Volume-Selective *z*-Shim Compensation. Mag Reson Med 2007;57:396-404.
16. Weiskopf N, Hutton C, Josephs O, Deichmann R. Optimal EPI parameters for reduction of susceptibility-induced BOLD sensitivity losses: A whole-brain analysis at 3 T and 1.5 T. NeuroImage 2006;33:493-504.
17. Kanwisher N, McDermott J, Chun MM. The Fusiform Face Area: A Module in Human Extrastriate Cortex Specialized for Face Perception. J Neurosci 1997;17:4302-4311.
18. Pessoa L, McKenna M, Gutierrez E, Ungerleider LG. Neural processing of emotional faces requires attention. Proc Natl Acad Sci USA 2002;99:11458-11463.
19. Lu H, Golay X, Pekar JJ, van Zijl PC. Intervoxel Heterogeneity of Event-Related Functional Magnetic Resonance Imaging Responses as a Function of T1 Weighting. NeuroImage 2002;17:943-955.
20. Bodurka J, Bandettini P. Physiological Noise Effects on the Flip Angle Selection in BOLD fMRI. 2009; Honolulu. p 221.

21. Ojemann JG, Akbudak E, Snyder AZ, McKinstry RC, Raichle ME, Conturo TE. Anatomic Localization and Quantitative Analysis of Gradient Refocused Echo-Planar fMRI Susceptibility Artifacts. NeuroImage 1997;6:156-167.
22. Maldjian JA, Laurienti PJ, Kraft RA, Burdette JH. An automated method for neuroanatomic and cytoarchitectonic atlas-based interrogation of fmri data sets. NeuroImage NeuroImage 2003;19:1233-1239.
23. Constable RT, Spencer DD. Composite Image Formation in $z$-Shimmed Functional MR Imaging. Mag Reson Med 1999;42:110-117.
24. Purdon PL, Weisskoff RM. Effect of Temporal Autocorrelation Due to Physiological Noise and Stimulus Paradigm on Voxel-Level False-Positive Rates in fMRI. Human Brain Mapping 1998;6:239-249.
25. Deichmann R, Gottfried JA, Hutton C, Turner R. Optimized EPI for fMRI studies of the orbitofrontal cortex. NeuroImage 2003;19:430-441.
26. Murphy K, Bodurka J, Bandettini PA. How long to scan? The relationship between fMRI temporal signal to noise and necessary scan duration. NeuroImage 2007;34:565-574.
27. Kruger G, Glover GH. The physiological noise in oxygen-sensitive magnetic resonance imaging. Mag Reson Med 2001;46:631-637.